\documentclass[12pt]{article}
\usepackage{epsfig,amsmath,amssymb,latexsym,graphicx}
\usepackage{setspace,fullpage}

\def\bib{\vskip12pt\par\noindent\hangindent=0.5 true cm\hangafter=1}

\newcommand{\by}{\mbox{\boldmath$y$}}

\newcommand{\bA}{\mbox{\boldmath$A$}}
\newcommand{\bx}{\mbox{\boldmath$x$}}
\newcommand{\bX}{\mbox{\boldmath$X$}}
\newcommand{\bY}{\mbox{\boldmath$Y$}}
\newcommand{\bC}{\mbox{\boldmath$C$}}
\newcommand{\bD}{\mbox{\boldmath$D$}}

\newcommand{\bH}{\mbox{\boldmath$H$}}
\newcommand{\bI}{\mbox{\boldmath$I$}}
\newcommand{\bR}{\mbox{\boldmath$R$}}
\newcommand{\bS}{\mbox{\boldmath$S$}}
\newcommand{\bT}{\mbox{\boldmath$T$}}
\newcommand{\bZ}{\mbox{\boldmath$Z$}}
\newcommand{\bw}{\mbox{\boldmath$w$}}

\newcommand{\bbeta}{\mbox{\boldmath$\beta$}}
\newcommand{\bfeta}{\mbox{\boldmath$\eta$}}

\newcommand{\bmu}{\mbox{\boldmath$\mu$}}

\newcommand{\bLambda}{\mbox{\boldmath$\Lambda$}}
\newcommand{\bdelta}{\mbox{\boldmath$\delta$}}
\newcommand{\btheta}{\mbox{\boldmath$\theta$}}

\begin{document}

{\baselineskip=18pt
{
\vspace*{10mm}
{\renewcommand{\thefootnote}{\fnsymbol{footnote}}
\begin{center}
{\LARGE
Approximating the marginal likelihood using copula } \\

 \vspace*{15mm}
 
{\Large David J. Nott, Robert Kohn and Mark Fielding\footnote{
David J. Nott is Associate Professor, Department of Statistics and Applied Probability,
National University of Singapore, Singapore 117546. (email
standj@nus.edu.sg).  Robert Kohn is Professor, Australian School of Business, 
University of New South Wales, Sydney 2052 Australia.  Mark Fielding is Postdoctoral Fellow, 
Department of Statistics and Applied Probability, National University of Singapore, 
Singapore 117546.  This work was partially supported
by Australian Research Council grant DP0667069.  The authors thank Professor Denzil Fiebig
for supplying the data of Section 7.}}

\end{center}

\doublespacing

\vspace*{4mm}
\begin{abstract}
Model selection is an important activity in modern
data analysis and the conventional Bayesian approach to this problem involves
calculation of marginal likelihoods for different models, together with
diagnostics which examine specific aspects of model fit.  
Calculating the marginal likelihood is a difficult computational problem.  
Our article proposes some extensions of the
Laplace approximation for this task that are related to copula models 
and which are easy to apply.  
Variations which can be used both with and without simulation from
the posterior distribution are considered, as well as use of the 
approximations with bridge sampling and in random effects models
with a large number of latent variables.  The use of a $t$-copula to obtain 
higher accuracy when multivariate dependence is not well captured by a Gaussian copula is also
discussed.  
\end{abstract}

\vspace*{5mm}\noindent

\section{Introduction}

In Bayesian inference computation of marginal likelihoods 
is essential for calculating posterior model probabilities 
and Bayes factors, fundamental quantities for model comparison in the Bayesian framework.  
If $M_1$ and $M_2$ are two models to be compared with respective parameters $\btheta_1$ and 
$\btheta_2$, priors $p(\btheta_1)$ and $p(\btheta_2)$ and likelihoods
$p(\by|\btheta_1)$ and $p(\by|\btheta_2)$ where $\by=(y_1,...,y_n)^T$ denotes
the data then if $p(M_1)$ and $p(M_2)$ denote the prior probabilities for $M_1$ and
$M_2$ then the ratio of their respective posterior probabilities is
\begin{eqnarray*}
  \frac{p(M_1|\by)}{p(M_2|\by)} & = & \frac{p(M_1)}{p(M_2)}\times \frac{p(\by|M_1)}{p(\by|M_2)}
\end{eqnarray*}
where $p(\by|M_j)=\int p(\btheta_j)p(\by|\btheta_j)d\btheta_j$ is the marginal likelihood
for model $M_j$ and the second term on the right side above is called the Bayes factor 
comparing $M_1$ to $M_2$.  
  
There are many suggestions for how to calculate the marginal likelihood.  
One of the simplest methods is the Laplace approximation.  We consider now a single model
$M$ with parameter $\btheta$ of dimension $p$, prior $p(\btheta)$ and likelihood
$p(\by|\btheta)$ with marginal likelihood
\begin{eqnarray}
  p(\by) & = & \int p(\btheta)p(\by|\btheta)d\btheta \label{mlike}.
\end{eqnarray}
Suppressing dependence on $\by$, write $f(\btheta)=p(\btheta)p(\by|\btheta)$ and $g(\btheta)=\log f(\btheta)$.  
Let $\hat{\btheta}$ be the mode of $g(\btheta)$ and $\bH$ be the negative Hessian
at the mode.  The Laplace approximation approximates $g(\btheta)$ by
$g(\hat{\btheta})-1/2(\btheta-\hat{\btheta})^T \bH (\btheta-\hat{\btheta})$. 
Substituting this into (\ref{mlike}) and integrating gives
\begin{eqnarray}
  p(\by) & \approx & (2\pi)^{p/2}|\bH|^{-1/2}f(\hat{\btheta})  \label{laplace}
\end{eqnarray}
where it can be shown that the error is of order $O(n^{-1})$.  The right side of (\ref{laplace})
is commonly used for approximating the marginal likelihood in Bayesian inference
but there has also been much
interest in the use of the Laplace approximation for calculation of posterior moments
in Bayesian applications (Tierney and Kadane, 1986).  O'Hagan and Forster (2004), Chapter 9, 
is a good summary of these applications and associated theory.  

There are many other suggested methods for computing the marginal likelihood which
are simulation based.  Raftery {\it et al.} (2007) and Newton and Raftery (1994) 
consider approaches based on using the so-called harmonic mean identity, an 
extension of which was discussed by Gelfand and Dey (1994).  Such an approach
can be unstable, although Raftery {\it et al.} (2007) suggest some possible solutions.  
Averaging the likelihood over parameters simulated from the prior is another possibility 
that directly uses the definition (\ref{mlike}), but since the prior is overdispersed
with respect to the likelihood this can be very inefficient requiring very large sample
sizes.  Several Markov chain Monte Carlo (MCMC) methods attempt to sample on the model
and parameter space jointly.  Carlin and Chib (1995) suggest an approach based
on simulating on a product space.  However this approach is hard to apply with a large
number of models and requires choice of some tuning parameters that make the 
method unsuited to routine use.  Green (1995) extends the Metropolis-Hastings algorithm
to situations involving model uncertainty and his trans-dimensional MCMC method is
the method of choice when there is a large number of models to be compared.  However, 
devising MCMC moves to jump between different models in this framework is an art
that requires problem specific insight.  Several methods for calculating the marginal
likelihood make use of the identity 
\begin{eqnarray}
 p(\by) & = & \frac{p(\btheta)p(\by|\btheta)}{p(\btheta|\by)}  \label{candform}
\end{eqnarray}
which holds for any value of $\btheta$ by a rearrangement of Bayes' rule.  To use this identity, 
simply observe that the numerator is easy to calculate at any $\btheta$ so that if we are
able to estimate the posterior distribution $p(\btheta|\by)$ at some point 
$\hat{\btheta}$ (usually some estimate of the mode) then we immediately
have an estimate of the marginal likelihood.  
The idea of estimating the marginal likelihood in this way is attributed
to Julian Besag by Raftery (1996).  
Chib (1995) considered use
of this identity in the case where a Gibbs' sampling MCMC algorithm is employed, 
although when the Gibbs updates consist of many blocks several different runs 
are needed to get the required density estimate.  Although there is a way of
avoiding multiple runs, this may not work well in high dimensions (see Chib and Jeliazkov, 2001, 
for further discussion).  Chib and Jeliazkov (2001) extended the method of Chib (1995) 
to the case of a general Metropolis-Hastings algorithm, although again if the Metropolis-Hastings
scheme updates the parameters in many different blocks the method can be tedious to apply.  
Mira and Nicholls (2004) show that the method of Chib and Jeliazkov (2001) is a special
case of the bridge estimator of Meng and Wong (1996) in the way that it estimates
certain conditional densities and they also discuss optimality of the
bridge sampling implementation.  de Valpine (2008) also considers
several further refinements of the approach.  Gelman and Meng (1998) extend
the bridge estimator of Meng and Wong (1986) to an approach they call path sampling -- it is related 
in statistics to a method for high-dimensional integration discussed by Ogata (1989)
and to previous work in statistical physics.  
Implementing path sampling can be quite computationally intensive, involving either several MCMC
runs for different target distributions or an MCMC run over a joint distribution including
an auxiliary variable.  
Friel and Pettitt (2008) provide one recent approach to the implementation of path sampling.  
Another recent novel approach to marginal likelihood calculation 
is given by Skilling (2006) although implementation
of this approach involves possibly difficult simulations from constrained distributions.    
Han and Carlin (2000) give a survey of MCMC based methods for computing the marginal
likelihood and suggest methods based on separate MCMC runs for different models such as those of Chib (1995) and
Chib and Jeliazkov (2001) as being easiest to use when the number of models to be compared is small.  

We restrict attention in what follows to methods which are easily employed
given a simulated sample from the posterior distribution.  
However, we also consider an extension of the Laplace approximation
which does not require simulation.  Ways of combining simulation and the Laplace
approximation for computing Bayes factors were considered in DiCiccio {\it et al.} (1997).  
They recommended for routine use a volume corrected version of the Laplace approximation, 
or for higher accuracy and where evaluations of the likelihood are inexpensive 
a Laplace approximation approach to attaining a near optimal implementation of 
bridge sampling.  
We discuss this last method later as the copula approximations we introduce
here can be improved in much the same way.  

In Section 2 we discuss generalizing the Laplace approximation by approximating
the posterior distribution with a Gaussian copula.  Density estimation using
the Gaussian copula is sensible since in many cases we expect the posterior
distribution to be close to normal, so that approximating the posterior with a flexible
class of densities which contains the Gaussian as a special case is attractive.  
We also discuss generalizations where the Gaussian copula is replaced by a $t$-copula.  
Copula approximations to posterior distributions have not been used very much for Bayesian
computation -- a notable exception is Reichert {\it et al.} (2002) who considered
their use in importance sampling schemes.  
In Section 3 we consider different ways to estimate the marginal distributions in
the copula approximation, both with and without simulation output.  
Section 4 discusses the Laplace bridge estimator which uses an initial estimate of
the marginal likelihood obtained by Laplace approximation in implementation of 
bridge sampling.  A similar estimator using our copula
framework is then considered.  Section 5 considers performance of our methods in
some simulated examples, Section 6 considers an example involving logistic regression, 
Section 7 considers a random effects heteroscedastic probit model for clustered
binary data with a large number of latent variables and
Section 8 concludes.  The copula approximations we describe work well across
the whole range of examples we consider, both real and simulated.  

\section{A copula Laplace approximation}

A Gaussian copula distribution for a continuous random
vector $\btheta=(\theta_1,...,\theta_p)^T$ is constructed 
from given marginal distributions $F_j(\theta_j)$ for $\theta_j$, $j=1,...,p$ and a correlation matrix for a 
latent Gaussian random vector.  In particular, suppose ${\bf Z}\sim N(0,\bLambda)$ where $\bLambda$ is
a correlation matrix.  Then if $\theta_j=F_j^{-1}(\Phi(Z_j))$ then $\theta_j$ has distribution $F_j$ since
$\Phi(Z_j)$ is uniform and transforming a uniform random variable by the inverse of $F_j$
gives a random variable with distribution function $F_j$.  Note that while the $\theta_j$ 
have given marginal distributions
$F_j$, they are are also correlated due to the correlation between the components of $\bZ$.  For
background on Gaussian copula and copula models more generally see Joe (1997). 
The density function of $\btheta$ is (see, for example, Song, 2000)
\begin{eqnarray}
q(\btheta) & = & |\bLambda|^{-1/2} \exp\left(\frac{1}{2}\bfeta(\btheta)^T(\bI-\bLambda^{-1})\bfeta(\btheta)\right)\prod_{j=1}^p f_j(\theta_j) \label{copula}
\end{eqnarray}
where $\bfeta=\bfeta(\btheta)=(\eta_1,...,\eta_p)^T$ with $\eta_j=\Phi^{-1}(F_j(\theta_j))$ and $f_j$ is 
the density function corresponding to $F_j$.    

Now suppose we are able to obtain a copula approximation to a posterior distribution $p(\btheta|\by)$ 
for a parameter $\btheta$, of the form (\ref{copula}).  
We discuss how to obtain such an approximation both with and without simulation later.  
Then given an estimate $\hat{\btheta}$ of the mode of $p(\btheta|\by)$ we can employ
the identity (\ref{candform}) and our copula density estimate at $\hat{\btheta}$ to obtain
the estimate
\begin{eqnarray}
  p(\by) & \approx & p(\hat{\btheta})p(\by|\hat{\btheta})\frac{|\bLambda|^{1/2}\exp\left(-\frac{1}{2}\bfeta(\hat{\btheta})^T(\bI-\bLambda^{-1})\bfeta(\hat{\btheta})\right)}{\prod_{j=1}^p f_j(\hat{\theta}_j)}.  \label{copulalaplace}
\end{eqnarray}
where $\bLambda$ is the correlation matrix and $f_j(\theta_j)$, $j=1,...,p$ are the marginal
densities in our copula approximation.  If $\hat{\btheta}$ is the componentwise posterior median 
then $\bfeta(\hat{\btheta})=0$ and we obtain
\begin{eqnarray}
p(\by) & \approx & \frac{p(\hat{\btheta})p(\by|\hat{\btheta})|\bLambda|^{1/2}}{\prod_{j=1}^p f_j(\hat{\theta}_j)}.
  \label{copulalaplace2}
\end{eqnarray}
This estimate reduces to the ordinary Laplace approximation if we consider the special case
where our Gaussian copula is a multivariate normal density estimate with mean the posterior 
mode and covariance matrix
given by the inverse of the negative Hessian of the log posterior at the mode.  

\section{Estimating the marginals}

To apply the approximation (\ref{copulalaplace}) we need a Gaussian copula approximation
to the posterior distribution.  We consider both analytic and simulation based methods for obtaining this, 
as well as an extension of the Gaussian copula approach which uses 
$t$-copula.  

\subsection{Analytic approach}

Write, as in Section 1, $f(\btheta)=p(\btheta)p(\by|\btheta)$, 
$g(\btheta)=\log f(\btheta)$ and $\bH=-g''(\hat{\btheta})$
for the matrix of negative second order partial derivatives of $g(\btheta)$ evaluated at the mode $\hat{\btheta}$.  
Decompose $\bH$ as $\bD\bC\bD^T$ where $\bC$ is a correlation matrix and $\bD=\mbox{diag}(d_j)$ and  
$\bH^{-1}=\bS\bA\bS^T$ where $\bA$ is a correlation matrix and $\bS=\mbox{diag}(s_j)$.  Now consider
a Gaussian copula density as an approximation to the posterior distribution, where the approximation
to the marginal posterior distribution for $\theta_j$ is 
$$f_j(\theta_j)=\frac{f(\hat{\btheta}+(\theta_j-\hat{\theta}_j){\bf e}_j)^{1/(d_j^2s_j^2)}}
{\int_{-\infty}^\infty f(\hat{\btheta}+(\theta_j-\hat{\theta}_j){\bf e}_j)^{1/(d_j^2s_j^2)}d\theta_j}$$
and ${\bf e}_j$ is a $p$-vector of zeros but with a one in the $j$th position 
and the copula correlation matrix is $\bA$.   

The intuition behind this density estimate is as follows.  We estimate the marginal for $\theta_j$
by considering a slice through the function $f(\btheta)$ with values of $\theta_i$, $i\neq j$, fixed
at their modal values (this is the function $f(\hat{\btheta}+(\theta_j-\hat{\theta}_j){\bf e}_j)$ 
and then overdisperse by raising this function to a power and normalizing.  
Note that if $f(\btheta)$ is proportional to a multivariate Gaussian, then
$f(\hat{\btheta}+(\theta_j-\hat{\theta}_j){\bf e}_j)$
is proportional to the conditional density of $\theta_j$ given that the other components are fixed at their modal
values.  This conditional distribution has as its mean the unconditional mean of $\theta_j$, and
the variance $1/s_j^2$.  Then raising this function to 
the power of $1/(d_j^2s_j^2)$ and normalizing maintains the mean while changing
the variance from $1/s_j^2$ to $d_j^2$, which is the unconditional variance for $\theta_j$ (in the
multivariate Gaussian case).  So this operation gives the correct marginal distribution when
$f(\btheta)$ is proportional to a multivariate Gaussian.  The approximation to the marginal is also exact in
the case of independence (where $d_j^2s_j^2=1$ and $\bA=\bI$).  Choice of the copula correlation
matrix as $\bA$ is also made to ensure that the approximation to the joint posterior is exact 
in the case of $p(\btheta|\by)$ being multivariate normal.   
The copula approximation is of interest in itself apart from the application to computing
marginal likelihoods.  In particular, for a
Gaussian copula expectations for low-dimensional marginal
distributions are easily calculated.  For instance, suppose that we want
to approximate for the $j$th component $\theta_j$ of $\btheta$ the posterior expectation
$E(h(\theta_j)|y)$.  Then this is easily obtained from our copula approximation as
$$\int h(\theta_j)g_j(\theta_j)d\theta_j$$
where $g_j(\theta_j)$ is the marginal for $\theta_j$.  This
expression is easily evaluated with one-dimensional numerical integration.  
The approximation is exact both in the Gaussian case and in the case where components
of the posterior are independent and it seems preferable to the simple normal approximation.  

\subsection{Simulation based approach}

An alternative and more accurate approach to approximating the posterior distribution
by a Gaussian copula involves using a simulation based method.  Suppose that we have
a sample $\btheta^{(1)},...,\btheta^{(s)}$ from the posterior distribution $p(\btheta|\by)$
obtained by some method such as MCMC.  Consider the estimate (\ref{copulalaplace2}) where
$\hat{\btheta}$ consists of the componentwise median.  
We estimate the quantities $f_j(\hat{\theta}_j)$ using kernel density estimates based
on the simulation output.  We note that Hsiao {\it et al.} (2004) have considered multivariate
kernel density estimation in conjunction with the formula (\ref{candform}) for
estimating the marginal likelihood but clearly this approach is limited to 
fairly low dimensional situations.  It only remains to specify 
how we obtain the correlation matrix
$\bLambda$ in our copula approximation to the posterior.  

Let $r_{jk}$ be the rank of $\theta_k^{(j)}$ among the values $\theta_k^{(i)}$, $i=1,...,s$.  
Define the $p$-vector $\bZ^{(j)}$ to have $k$th component
$Z_k^{(j)}=\Phi^{-1}((r_{jk}-0.5)/s)$ where $\Phi$
denotes the standard normal distribution function.  We obtain $\bLambda$ as the estimated 
correlation matrix of $\bZ^{(1)},...,\bZ^{(s)}$, which we obtain by the robust method
of Rosseuw and Van Zomeren (1990) rather than using the sample correlation matrix, 
similar to Di Ciccio {\it et al.} (1997) in their implementation
of a simulation based Laplace approximation.  
Roughly speaking, the above construction estimates
the marginal distribution for a copula with the empirical distribution function and
then transforms to the latent Gaussian variables assumed in the copula construction
to obtain an estimate of the copula correlation.  

We can extend our Gaussian copula approximation to a $t$-copula.  Laplace-type approximations
using the multivariate $t$-distribution have been considered previously (Leonard, Hsu and Ritter, 1994).  A $t$-copula distribution with $\nu$ degrees of freedom for
a continuous random vector $\btheta=(\theta_1,...,\theta_p)^T$ with marginals
$F_1(\theta_1),...,F_p(\theta_p)$ has density 
$$q(\btheta)=\frac{f_p(\bfeta(\btheta); 0,\bLambda, \nu)}{\prod_{j=1}^p f_1(\eta_j(\theta_j); 0,1,\nu)}\prod_{j=1}^p f_j(\theta_j)$$
where $f_k(\bfeta;0,\bLambda,\nu)$ is the $k$-dimensional multivariate $t$-density
with mean $0$, scale $\bLambda$ and degrees of freedom $\nu$, $\bfeta(\btheta)=(\eta_1,...,\eta_p)^T$ with $\eta_j=\eta_j(\theta_j)=F_1^{-1}(F_j(\theta_j);0,1,\nu))$ where $F_k(\bfeta;0,\bLambda,\nu)$ is the distribution function for $f_k(\bfeta;0,\bLambda,\nu)$, and $f_j(\theta_j)$ is
the density for $F_j(\theta_j)$.  If we have an approximation to the posterior
distribution of this form we can again obtain an estimate of the marginal likelihood
based on the estimated posterior density at some value $\hat{\btheta}$.  Taking again 
$\hat{\btheta}$ as the componentwise median, we obtain
$$p(y)\approx \frac{p(\hat{\btheta})p(\by|\hat{\btheta})|\bLambda|^{1/2}}{\prod_{j=1}^p f_j(\hat{\theta}_j)} \frac{\Gamma\left(\frac{\nu+1}{2}\right)^p}{\Gamma\left(\frac{\nu+p}{2}\right)\Gamma\left(\frac{\nu}{2}\right)^{p-1}}.$$
Of course, as $\nu\rightarrow\infty$ this reduces to our former approximation based
on the Gaussian copula.  Once more we need some simple way to estimate $\bLambda$ and
in this case $\nu$ from simulation output to apply this formula.  Let $r_{jk}$ be
the rank of $\theta_k^{(j)}$ among the values $\theta_k^{(i)}$, $i=1,...,s$.  
For fixed degrees of freedom $\nu$ let $T_k^{(j)}=F_1^{-1}((r_{jk}-0.5)/s; 0,1,\nu)$, 
$\bT^{(j)}=(T_1^{(j)},...,T_p^{(j)})^T$ and let $\bLambda(\nu)$ be obtained as
the maximum likelihood estimator of the correlation matrix assuming the 
$\bT^{(j)}$ are independent and identically distributed from a multivariate $t$-distribution with mean $0$, scale
$\bLambda$ and degrees of freedom $\nu$.  On the copula scale one can consider the data 
$\bR^{(j)}$ with $R_k^{(j)}=(r_{jk}-0.5)/s$ and assuming the $\bR^{(j)}$ are independent and
identically distributed from the density
$$\frac{f_p(F_1^{-1}(r_1;0,1,\nu),...,F_p^{-1}(r_p;0,1,\nu);0,\bLambda(\nu),\nu)}
{\prod_{j=1}^p f_1(F_1^{-1}(r_i;0,1,\nu);0,1,\nu)}$$
obtain a maximum likelihood estimator for $\nu$ numerically by a grid search.  

\section{Laplace bridge estimator}

DiCiccio {\it et al.} (1997) find that combining Laplace approximation and the bridge
estimator of Meng and Wong (1996) is very effective in improving accuracy.  
In its most general form the bridge estimator can estimate a ratio of marginal
likelihoods but here we just consider a special case where interest centres on
calculation of a single marginal likelihood.  We want to calculate the normalizing 
constant (marginal likelihood) $p(\by)$ in
$p(\btheta|\by) \propto p(\btheta)p(\by|\btheta)$.  
Suppose we have some density $r(\btheta)$ (where in this discussion there is no unknown
normalizing constant for $r$) and let $t(\btheta)$ be any function of $\btheta$ 
such that 
$$0<\left| \int t(\btheta)r(\btheta)p(\btheta)p(\by|\btheta) d\btheta \right|<\infty.$$
Then it is easily shown that 
$$p(\by)=\frac{\int p(\btheta)p(\by|\btheta)t(\btheta)r(\btheta)d\btheta}
{\int r(\btheta)t(\btheta)p(\btheta|\by)d\btheta}.$$
If we have a sample $\btheta^{(1)},...,\btheta^{(s)}$ from $p(\btheta|\by)$ and a sample
$\tilde{\btheta}^{(1)},...,\tilde{\btheta}^{(S)}$ from $r(\btheta)$, then we have
$$p(\by)\approx \frac{\frac{1}{S}\sum_{i=1}^S t(\tilde{\btheta}^{(i)})p(\tilde{\btheta}^{(i)})p(\by|\tilde{\btheta}^{(i)})}
{\frac{1}{s}\sum_{i=1}^s t(\btheta^{(i)}) r(\btheta^{(i)})}.$$
Meng and Wong (1996) show that the optimal choice of the function $t(\btheta)$ is
\begin{eqnarray}
  \left\{ s \frac{p(\btheta)p(\by|\btheta)}{p(\by)}+S r(\btheta) \right\}^{-1}. \label{optt}
\end{eqnarray}
Actually, (\ref{optt}) is the optimal choice when the generated samples from both
$p(\btheta|\by)$ and $r(\btheta)$ are independent.  For the samples from $p(\btheta|\by)$
which are usually generated via MCMC this is usually not the case and an adjustment could 
be made to account for the typically positive correlation, although we have not done this here.  
Note also that (\ref{optt}) involves $p(\by)$, which is unknown.  
It is possible to implement an iterative version of bridge sampling where $p(\by)$ is
successively refined in (\ref{optt}) (Meng and Wong, 1996).  The Laplace bridge estimator
simply uses the Laplace approximation to estimate $p(\by)$ in (\ref{optt}) for the purpose of
determining a $t(\btheta)$ for implementation of bridge sampling, and uses
the usual normal approximation for $r(\btheta)$.  We can similarly suggest estimating
$p(\by)$ with our Gaussian copula approach, and using our Gaussian copula approximation
to the posterior for $r(\btheta)$ with kernel density estimates based on the simulation
output for the marginals.  Note that simulating directly from a Gaussian copula is
straightforward.  

\section{Simulation studies}

To evaluate the accuracy of our methods we consider their use for calculating
the normalizing constant for some known density functions.  Of course, the normalizing
constant is known to be one here so accuracy of the approximations is easily assessed.  
We consider the multivariate skew $t$ distribution of Branco and Dey (2001).  
This is a convenient distribution to use since it can accommodate both skewness
and heavy tails, it is easy to simulate from non-iteratively and its density function
is easy to calculate.  Branco and Dey (2001), p. 105, consider a generalized multivariate 
skew $t$ distribution but here we just consider the special case of their multivariate
skew $t$.  For $\bY=(Y_1,...,Y_k)^T$ the density is
\begin{eqnarray}
  f_Y(\by) & = & 2 f_k(\by;\bmu,\bLambda,\nu)F_1\left(\frac{\bdelta^T \bLambda^{-1} (\by-\bmu)}
  {\sqrt{1-\bdelta^T\bLambda^{-1} \bdelta}}\sqrt{\frac{\nu+k}{\nu+(\by-\bmu)^T\bLambda^{-1}(\by-\bmu)}};0,1,\nu+k\right) \label{mvskewt}
\end{eqnarray}
where as before $f_k(\by;\bmu,\bLambda,\nu)$ is the $k$-dimensional 
multivariate $t$ distribution with
mean $\bmu$, scale matrix $\bLambda$ and $\nu$ degrees of freedom, 
$F_k(\by;\mu,\bLambda,\nu)$ is
the corresponding density function, and $\bdelta=(\delta_1,...,\delta_k)^T$ is 
a vector of skewness parameters.  Obviously with $\bdelta=0$ we obtain the ordinary
multivariate $t$ distribution.  For our simulations we choose $\bmu=0$, $\bLambda=\bI$ 
and $\bdelta$ of the form $(\delta_1,0,...,0)^T$.  
Choosing $\bLambda$ and $\bdelta$ in this way means that only the first component of
$\bY$ is skewed, with $\delta_1$ controlling skewness, $\delta_1>0$ giving positive skewness
and $\delta_1<0$ giving negative skewness.  The random vector $\bY$ with density (\ref{mvskewt})
may be constructed in the following way.  Let $\bZ=[X_0 \, \bX^T]^T$ be a $(k+1)$-dimensional
multivariate $t$ distributed random vector with $X_0$ a scalar and $\bX$ a $k$-vector, 
$$\bmu^*=(0,\bmu^T)^T \;\;\;\;\; \bLambda^*=\left[\begin{array}{ll} 1 & \bdelta^T \\
 \bdelta & \bLambda \end{array}\right]$$
where $\bmu^*$ and $\bLambda^*$ are partitioned in the same way as $\bZ$.  Then $\bY$ has
the distribution of $\bX|X_0>0$.  Note that this construction gives a simple
non-iterative way of simulating from this distribution.  

Our simulations investigated the effects of heavy tails (through $\nu$), skewness
(through $\delta_1$) and dimensionality on the accuracy of the method we have
discussed.  In particular, for each of our methods
we considered every combination of $\nu=3,10$, $\delta_1=0,0.5,0.99$ and
$k=2,5,10$ dimensions.  The methods we compare are
the ordinary Laplace approximation (L1), a Laplace approximation where we use
the componentwise posterior median and minimum volume ellipsoid covariance estimation method of
Rosseuw and Van Zomeren (1990) from simulation output rather than
the mode and negative inverse Hessian (L2), our copula Laplace approximation without simulation 
(CL1), our copula approximation with simulation (CL2), the $t$-copula approximation
(TC), the Laplace bridge estimator (LB) and the copula bridge estimator
(CLB).  For the methods based on simulation, we used 10,000 replications.  
For method L2, the use of the componentwise median and minimum volume ellipsoid methods
for estimating the mean and covariance were discussed in DiCiccio {\it et al.} (1997)
and found to work well in high dimensions.  
For the two variants of bridge estimation we also used 10,000 simulations from
$r(\btheta)$, and for the copula bridge estimator
we estimated the marginals using a kernel density estimator (we used the default implementation
of the {\tt density} function in the {\tt stats} package of R, R core development team, 2005).  
For the simulation based methods we report average values obtained over
50 simulation replicates, with the standard deviation over replicates in brackets.
Of course there are ways to approximate standard errors based on a single replicate (de Valpine, 2008, 
for example) but we have chosen not to do this here for the purposes of our simulation studies.  
In practice such methods are of course very important.    
In our tables we have reported the estimated value of the log of the normalizing constant (true value 0) rather than
the normalizing constant itself.  
The results are shown in Tables 1-3.  The main conclusions which emerge are that the copula
approximations improve over the respective Laplace approximations for the variants both
with and without simulation.  Generally performance of all methods deteriorates with
higher dimension and heavier tails, as might be expected.  Perhaps not intuitively 
for some of the methods performance improves with increasing skewness.  Both variants
of bridge estimation work well, but the copula bridge method seems to improve over
the Laplace bridge method in the 10-dimensional case with skewness, with a smaller standard
deviation over replicates.  
The $t$-copula works extremely well, but
perhaps this is not surprising given that the test function is constructed from 
a generalization of the multivariate $t$-distribution.  In the $t$-copula, 
the estimate of the degrees of freedom was chosen from a grid including
integer values $1$ to $10$, $15$, $20$ and $50$.  In our later real examples the
$t$-copula approximation fares less well than in these simulations.  

\section{Low birthweight example}

For a real example we consider calculation of marginal likelihoods for model comparison 
in a regression with binary response.  In particular, we consider the low birth weight data reported by Hosmer and
Lemeshow (1989) which are concerned with 189 births at a US hospital in a study where it was
desired to found out which predictors of low birthweight were important in the hospital where the
study was carried out.  The binary response is an indicator for birthweight being less than 2.5 kg.  
After transforming the predictors as described in Venables and Ripley (2002) there are ten covariates, 
two continuous and eight binary predictors.  These covariates are shown in Table \ref{birthwtpred}.  
For our analysis of the data, we consider generalized linear models with logit and robit links (with three degrees
of freedom for the robit link), using
the default prior for logistic regression given by Gelman {\it et al.} (2008) on coefficients for both
the choices of link function.  See Gelman and Hill (2007) pp. 124-125 for a brief introduction to 
robit regression.  Venables and Ripley (2002) consider model selection for this example and the logit
link using a stepwise approach.  They also consider the inclusion of second order interaction
terms.  The final model they choose includes all the predictors in Table \ref{birthwtpred} as main
effects except for the indicators for race, as well as interaction terms {age*ftv1}, 
{age*ftv2+} and {smoke*ui}.  Here we consider a direct
comparison of the model including all the original predictors as main effects with
the final model of Venables and Ripley (2002) for both the logit and robit links.  Note that the
comparisons between links here are non-nested and not easily done via traditional hypothesis testing approaches.  
Venables and Ripley (2002) also consider examining the adequacy of 
their linear model with second order interactions by expanding to a generalized additive model
with smooth terms for the covariates age, age*ftv1, age*ftv2+ and lwt.  
We consider a similar model here, but we simply use second order polynomials for representing
the additive smooth terms which should be adequate for the purposes of model checking.  
For these three different models and the two different choices of link function we calculated
the log marginal likelihood using the same methods considered in our simulation study.  
We also considered the same comparisons for a random sample from the original data set of size $n=50$  
to show how the accuracy of the approximations is affected by sample
size.  For generating MCMC iterates we used a Metropolis-Hastings scheme with normal random
walk proposal with the covariance based on the Hessian of the log posterior at the mode. 
The results of these comparisons are shown in Tables \ref{bwtresults1} and \ref{bwtresults2}.  
Also shown in the table is a ``gold standard" value (GS) for each case obtained by Laplace
bridge with $s=S=100,000$.    
Using this value for comparison, we obtain a similar
picture of the performance of the respective methods to that obtained from the simulation
study.  All methods are remarkably accurate for the full data set.  In the small sample
setting of the randomly chosen subset, 
the copula approximations improve over their simpler Laplace approximation variants, 
and bridge sampling works well with the copula approach showing less variability than the Laplace bridge
for the highest-dimensional case.  For the $t$-copula, we estimated the
degrees of freedom choosing from a grid of integer values where the maximum
value is $50$ -- generally the largest value of $50$ was chosen in nearly
every case, so that performance is generally similar but slightly inferior to
the Gaussian copula approximation.

\section{A high-dimensional example}

Our last example concerns a complex random effects model for a dataset concerned
with stated preferences of Australian women on whether or not to have a papsmear
test (Fiebig and Hall, 2005).  There are 79 women in the study and each
is presented with 32 different scenarios.  The response is an indicator for whether
the women would undertake a papsmear test so there are 32 repeated binary observations 
on each of the 79 women.  We consider the following random effects
heteroscedastic probit model which was considered in Gu {\it et al.} (2008) and
analyzed using a Bayesian approach.  Following the notation of Gu {\it et al.} (2008)
and letting $i=1,...,79$ index the different women or clusters, and $j=1,...,32$ index
observations within clusters, the binary observation $y_{ij}$ is considered to
arise from a continuous latent variable $y_{ij}^*$ by
$$y_{ij}=\left\{\begin{array}{ll}
  1 & \mbox{if $y_{ij}^*>0$} \\
  0 & \mbox{otherwise.}
  \end{array}\right.$$
Similar latent variable formulations are often used in Bayesian analyses of
simple probit models (Albert and Chib, 1993).  The $y_{ij}^*$ follow
the model
$$y_{ij}^*=\bx_{ij}\bbeta+\mu_i+\nu_{ij}$$
where $\mu_i$ is a subject or cluster specific random effect, $\bx_{ij}$
is a vector of covariates, $\bbeta$ is an unknown vector of regression
coefficients and $\nu_{ij}\sim N(0,\sigma_{ij}^2)$ with 
$\sigma_{ij}^2=\exp(\bw_{ij} \bdelta)$ 
where $\bw_{ij}$ is a vector of covariates
(often $\bx_{ij}=\bw_{ij})$ and $\bdelta$ is a vector of unknown
coefficients.  For identifiability an intercept should not be
included in $\bw_{ij}$.  The covariates used to define $\bx_{ij}$
and $\bw_{ij}$ in this example are shown in Table \ref{papsmear}.  Gu {\it et al.} (2008)
use the following priors for $\bbeta$, $\bdelta$ and $\sigma_\mu^2$.  First, 
$$\bbeta|\bdelta \sim N(0,c_\beta (\tilde{\bX}^T\tilde{\bX})^{-1})$$
where $c_\beta$ is set to the total number of observations ($32\times 79$ here)
and $\tilde{\bX}=\bD(\bdelta)^{-1}\bX$ where $\bX=(\bx_{11}^T,...,\bx_{1,32}^T,...,\bx_{79,32}^T)^T$ and 
$$\bD(\bdelta)=\mbox{diag}\left(\exp\left(\frac{\bw_{11}\bdelta}{2}\right),...,
\exp\left(\frac{\bw_{1,32}\bdelta}{2}\right),...,\exp\left(\frac{\bw_{79,32}\bdelta}{2}\right)\right)^T.$$  
Then $\bdelta\sim N(0,c_\delta \bI)$ and $\sigma_\delta^2,c_\delta\sim \mbox{IG}(a,b)$ 
independently where $a=1+10^{-10}$, $b=1+10^{-5}$ and IG denotes the
inverse gamma distribution.  An efficient MCMC sampling scheme can be developed
with $\bbeta$ and $\bmu=(\mu_1,...,\mu_{79})^T$ updated as a single block
with a Gibbs sampling step, $\bdelta$ udpated using a Metropolis-Hastings
step and $\sigma_\delta^2$ and $c_\delta$ updated with Gibbs sampling
steps.  See Gu {\it et al.} (2008) for details.  
If we set $\bdelta=0$ in this model, this results in a homoscedastic random
effects probit model and it is of some interest to compare this model with the
full model.  See Gu {\it et al.} (2008) for references and discussion.  

This is a challenging example because of the presence of the latent variables $y_{ij}^*$ and 
$\bmu$.  Our approach effectively integrates out the latent
variables which is important since otherwise we obtain a very high-dimensional
problem.  We will apply the formula (\ref{candform}) for estimating the marginal
likelihood with $\btheta=(\bdelta^T,\bbeta^T,\sigma_\mu^2,c_\delta)^T$.  
Note that it is difficult to apply bridge sampling with $\bmu$ integrated
out as this requires evaluating $p(y|\btheta)$ for a large number
of different values of $\btheta$.  To apply our approach we need to be able to
estimate $p(\by|\btheta)$ at a single value $\btheta^*$.  
As before, assume that we have an MCMC sample from $p(\btheta,\by^*,\bmu|\by)$.  
We can use for $\btheta^*=({\bdelta^*}^T,{\bbeta^*}^T,{\sigma_\mu^2}^*,c_\delta^*)^T$
the componentwise posterior median, say.  Then to estimate $p(\by|\btheta^*)$ 
we can simulate values $\bmu^{(1)},...,\bmu^{(s)}$ from $p(\bmu|{\sigma_\mu^2}^*)$ and
compute 
$$\frac{1}{s}\sum_{i=1}^s p(\by|\btheta^*,\mu^{(i)}).$$
To use (\ref{candform}) to estimate $p(\by)$, it only remains to estimate
$p(\btheta^*|\by)$.  With our copula approach, we can do this directly by fitting
a Gaussian copula model to the simulation output.  We also consider a simple normal
density estimate, which is similar to the Laplace-Metropolis estimator of
Lewis and Raftery (1997).  For comparison, we implement the computationally
intensive but also more accurate method of Chib and Jeliazkov (2001) which
can be applied with latent variable models such as the one considered here.  
Write 
\begin{eqnarray}
 p(\btheta^*|\by) & = & p(\bdelta^*|\by)p(\bbeta^*|\bdelta^*,\by)p({\sigma_\mu^2}^*|\bbeta^*,\bdelta^*,\by)p(c_\delta^*|\bbeta^*,\bdelta^*,{\sigma_\mu^2}^*,\by).
   \label{CJ}
\end{eqnarray}
In the present context, we use the approach of Chib and Jeliazkov (2001) to
estimate each of the terms on the right hand side of (\ref{CJ}).  This requires
separate runs for the different blocks of parameters in the decomposition 
(see Chib and Jeliakov, 2001, for further discussion).  The terms for
$\bbeta$, $\sigma_{\mu}^2$ and $c_{\delta}$ are relatively easily handled as
full conditionals are available for these parameters, but $\bdelta$ is updated
by a Metropolis-Hastings step.  It is of interest to see whether the rather tedious but 
accurate multiple runs approach of Chib and Jeliazkov (2001) results in
very similar results to an approximation which requires less coding effort and
computation time.  Table \ref{CJresults} shows the results for the approach
of Chib and Jeliazkov (CJ), copula approximation (CL) and normal approximation
(L).  The log marginal likelihood is estimated for both heteroscedastic and 
homoscedastic models.  

The copula approximations work well for much less
computational effort.  The additional computational effort for the coupula approximation
is essentially negligible once the MCMC run for the full model is obtained
whereas CJ requires 2 additional reduced MCMC runs where first $\bdelta^*$
and then $\bbeta^*$ and $\bdelta^*$ are held fixed.  In the table
as before we report a mean and standard deviation (bracketed) across
50 simulation replicates for each of the methods.  We also tried the $t$-copula
approximation but this was similar but slightly inferior to the Gaussian
copula approximation.  

We can envisage a role for our copula approximations in conjunction with
the CJ approach and similar approaches in high-dimensional situations.  One could
use a copula approximation for some blocks of parameters in estimating the 
conditional distribution in (\ref{CJ}).  When it is natural to use
a large number of small blocks in the MCMC scheme the method of CJ may be very
tedious to apply so grouping some small blocks together and applying a copula
approximation while dealing with the remaining blocks using the CJ approach
(for instance for blocks where the full conditional is available) is potentially
attractive.  The greater accuracy of the copula approximation compared to
the normal approximation would allow the consideration of a larger number
of smaller blocks. 

\section{Discussion and Conclusions}

With large datasets becoming increasingly common in statistical applications 
there has been recent renewed interest in fast deterministic approximations like the
Laplace approximation as an alternative to Monte Carlo methods or to improve
the implementation of Monte Carlo methods in certain problems.  Among the methods
we have considered, the copula approximations are the ones that work well across
the whole range of real and simulated examples that we have discussed, and the
copula methods usually improve on their simpler Laplace type variants both
with and without simulation.  
We believe our methods have great potential to be used both by themselves, in combination with
other methods, and even in conjunction with MCMC algorithms where there is a need
for better proposal distributions.  
Investigation of these applications is continuing.  

\section*{References}

\bib
Albert, J. and Chib, S. (1993).  Bayesian analysis of binary and polychotomous 
response data.  {\it Journal of the American Statistical Association}, 
88, 669--79.  

\bib
Branco, M.D. and Dey, D.K. (2001).  
A general class of multivariate skew-elliptical distributions.  
{\it J. Multiv. Anal.}, 79, 99--113.  

\bib
Carlin, B. and Chib, S. (1995).  Bayesian model choice via Markov
chain Monte Carlo.  {\it J. Roy. Statist. Soc. B}, 57, 473--484.  

\bib
Chib, S. (1995).  Marginal likelihood from the Gibbs output.  
{\it J. Amer. Statist. Assoc.}, 90, 1313--1321.  

\bib
Chib, S. and Jeliazkov, I. (2001).  Marginal likelihood from the
Metropolis-Hastings output.  {\it J. Amer. Statist. Assoc.}, 
96, 270--281.  

\bib
de Valpine, P. (2008).  Improved estimation of normalizing 
constants from Markov chain Monte Carlo output.  
{\it J. Comp. Graph. Statist.}, 17, 333--351.  

\bib
DiCiccio, T.J., Kass, R.E., Raftery, A.E. and Wasserman, L. (1997).  
Computing Bayes factors by combining simulation and asymptotic 
approximation.  {\it J. Amer. Statist. Assoc.}, 92, 903--915.  

\bib
Fiebig, D.G. and Hall, J. (2005).  Discrete choice experiments 
in the analysis of health policy.  {\it Productivity Commission Conference, 
November 2005:  Quantitative Tools for Microeconomic Policy Analysis, Chapter 6}, 
119--136.  

\bib
Friel, N. and Pettitt, A.N. (2008).  Marginal likelihood estimation
via power posteriors.  {\it J. Roy. Statist. Soc. B}, to appear.  

\bib
Gelfand, A.E. and Dey, D.K. (1994).  Bayesian model choice:  Asymptotics
and exact calculations.  {\it J. Roy. Statist. Soc. B}, 56, 501--514.  

\bib
Gelman, A. and Meng, X.-L. (1998).  Simulating normalizing constants:  
From importance sampling to bridge sampling to path sampling.  
{\it Statistical Science}, 13, 163--185.  

\bib
Gelman, A. and Hill, J. (2007).  Data analysis using regression and multilevel/hierarchical
models.  Cambridge:  Cambridge University Press.  

\bib
Gelman, A., Jakulin, A., Pittau, M.G., and Su, Y.-S. (2008).  
A default prior distribution for logistic and other regression models.  
Technical report, 
available at \\
\verb+http://www.stat.columbia.edu/~gelman/research/unpublished/priors7.pdf+

\bib
Green, P.J. (1995).  Reversible jump Markov chain Monte Carlo computation
and Bayesian model determination.  {\it Biometrika}, 82, 711--732.  

\bib
Gu, Y., Fiebig, D.G., Cripps, E.J. and Kohn, R. (2008).  
Bayesian estimation of a random effects heteroscedastic probit model.  
Preprint available at
\verb+http://ssrn.com/abstract=1260140+

\bib
Hills, S.E. and Smith, A.F.M. (1993).  
Diagnostic plots for improved parametrization in Bayesian inference.  
{\it Biometrika}, 80, 61--74.  

\bib
Hosmer, D.W. and Lemeshow, S. (1989).  Applied Logistic Regression. New York: Wiley.

\bib
Hsiao, C.K., Huang, S.-Y., Chang, C.-W. (2004).  
Bayesian marginal inference via candidate's formula.  
{\it Statistics and Computing}, 14, 59--66.

\bib
Joe, H. (1997).  Multivariate models and dependence concepts.  London:  Chapman and Hall.  

\bib
Leonard, T, Hsu, J.S.J. and Ritter, C. (1994).  
The Laplacian $t$-approximation in Bayesian inference.  
{\it Statistica Sinica}, 4, 127-142 

\bib
Lewis, S.M. and Raftery, A.E. (1997).  
Estimating Bayes factors via posterior simulation with the Laplace-Metropolis
estimator.  {\it Journal of the American Statistical Association}, 
92, 648--655.  

\bib
Meng, X.-L., and Wong, W. (1996).  Simulating normalizing constants via
a simple identity:  A theoretical exploration.  {\it Statistica Sinica}, 
6, 831--860.  

\bib
Mira, A. and Nicholls, G. (2004).  Bridge estimation of the probability
density at a point.  {\it Statistica Sinica}, 14, 603--612.

\bib
Newton, M.A. and Raftery, A.E. (1994).  Approximate Bayesian
inference by the weighted likelihood bootstrap (with discussion).  
{\it J. Roy. Statist. Soc. B}, 56, 3--48.  

\bib
Ogata, Y. (1989).  A Monte Carlo method for high dimensional integration.  
{\it Numer. Math.}, 55, 137--157.  

\bib
O'Hagan, A. and Forster, J. (2004).  {\it Kendall's Advanced Theory of Statistics Volume 2B:  Bayesian
Inference (Second Edition)}.  London:  Arnold.    

\bib
R Development Core Team (2005). R: A language and environment for
statistical computing. R Foundation for Statistical Computing,
Vienna, Austria. ISBN 3-900051-07-0, URL http://www.R-project.org.

\bib
Raftery, A.E. (1996).  Hypothesis testing and model selection.  In:  W.R. Gilks, D.J. Spiegelhalter
and S. Richardson (Eds.),  {\it Markov chain Monte Carlo in Practice}, 
pp. 163-188.  London:  Chapman and Hall.

\bib
Raftery, A.E., Newton, M.A., Satagopan, J.M. and Krivitsky, P.N. (2007).  
Estimating the integrated likelihood via posterior simulation using the harmonic
mean identity (with discussion).  In:  J.M. Bernardo, M.J. Bayarri, J.O. Berger, A.P. Dawid, D. Heckerman, 
A.F.M. Smith and M. West (Eds.), Bayesian Statistics 8, pp. 1--45. 
Oxford:   Oxford University Press.  

\bib
Reichert, P., Schervish, M. and Small, M.J. (2002).  
An Efficient Sampling Technique for Bayesian Inference With Computationally Demanding Models.  
{\it Technometrics}, 44, 318--327.  

\bib
Rosseuw, P.J. and Van Zomeren, B.C. (1990).  
Unmasking multivariate outliers and leverage points (with discussion).  
{\it J. Amer. Statist. Assoc.}, 85, 633--651.

\bib
Skilling, J. (2006).  Nested sampling for general Bayesian computation.  
{\it Bayesian Analysis}, 1, 833--860.  

\bib
Song, X.-K. P. (2000).  Multivariate dipsersion models generated from Gaussian copula.  
{\it Scand. J. Statist.}, {\bf 27}, 305--320.  

\bib
Tierney, L. and Kadane, J.B. (1986).  Accurate approximations for posterior moments and
marginal densities.  {\it J. Amer. Statist. Assoc.}, {\bf 81}, 82--86.  

\bib
Venables, W.N. and Ripley, B.D. (2002).  Modern Applied Statistics with S.  Fourth Edition.  
New York:  Springer. 

\begin{table}[p]
\caption{\label{skewtresults1}
Methods L1, L2, CL1, CL2, TC, LB and CLB applied to integrate multivariate
skew t densities in $2$ dimensions with $3$ and $10$ degrees of freedom and
zero, moderate and extreme skewness. The estimated log of the integral of the
density (true value $0$) is reported. } 
\begin{center}
\begin{tabular}{c|c|ccc} 
\multicolumn{5}{c}{} \\ \hline
Degrees of  & Method & \multicolumn{3}{c}{Skewness}  \\ \cline{3-5}
freedom       &               &       $\delta_1=0$  &         $\delta_1=0.5$ &       $\delta_1=0.99$ \\ \cline{3-5}
   3             & L1           &     -0.51             &    -0.51              &     -0.60             \\
                 & L2           &     0.09 (0.02)       &    0.07 (0.02)        &     -0.30 (0.02)      \\
                 & CL1          &     -0.16             &    -0.17              &     -0.17             \\
                 & CL2          &     0.20 (0.02)       &    0.18 (0.03)        &     0.09 (0.03)       \\
                 & TC           &     0.04 (0.03)       &    0.03 (0.02)        &     -0.01 (0.02)       \\
                 & LB           &     0.00 (0.01)       &    0.00 (0.01)        &     0.00 (0.01)       \\
                 & CLB          &     0.00 (0.00)       &    0.00 (0.00)        &     0.00 (0.01)       \\     \hline
   10            & L1           &     -0.18             &    -0.18              &     -0.34             \\
                 & L2           &     -0.03 (0.01)      &    -0.04 (0.02)       &     -0.27 (0.02)      \\
                 & CL1          &     -0.05             &    -0.05              &     -0.05             \\
                 & CL2          &     0.07 (0.03)       &    0.07 (0.03)        &     0.04 (0.03)       \\
                 & TC           &     0.02 (0.03)       &    0.02 (0.03)        &     0.01 (0.03)       \\
                 & LB           &     0.00 (0.00)       &    0.00 (0.00)        &     0.00 (0.01)       \\
                 & CLB          &     0.00 (0.00)       &    0.00 (0.00)        &     0.00 (0.00)       \\     \hline  
\end{tabular}
\end{center}
\end{table}             

\begin{table}[p]
\caption{\label{skewtresults2}
Methods L1, L2, CL1, CL2, TC, LB and CLB applied to integrate multivariate
skew t densities in $5$ dimensions with $3$ and $10$ degrees of freedom and
zero, moderate and extreme skewness. The estimated log of the integral of the
density (true value $0$) is reported. } 
\begin{center}
\begin{tabular}{c|c|ccc}
\multicolumn{5}{c}{} \\ \hline
Degrees of  & Method & \multicolumn{3}{c}{Skewness}  \\ \cline{3-5}
freedom       &               &       $\delta_1=0$  &         $\delta_1=0.5$ &       $\delta_1=0.99$ \\ \cline{3-5}
   3             & L1           &     -1.55             &    -1.55              &     -1.69             \\
                 & L2           &     1.08 (0.03)       &    1.02 (0.03)        &     0.50 (0.04)      \\
                 & CL1          &     -1.04             &    -1.05              &     -1.06             \\
                 & CL2          &     0.66 (0.04)       &    0.60 (0.05)        &     0.32 (0.06)       \\
                 & TC           &     0.08 (0.05)       &    0.04 (0.05)        &    -0.01 (0.06)       \\
                 & LB           &     0.00 (0.01)       &    0.00 (0.01)        &     0.00 (0.02)       \\
                 & CLB          &     0.00 (0.01)      &    0.00 (0.01)       &     0.00 (0.01)       \\     \hline
   10            & L1           &     -0.68             &    -0.68              &     -0.85             \\
                 & L2           &     -0.31 (0.03)      &    0.30 (0.03)       &      0.08 (0.04)      \\
                 & CL1          &     -0.42             &    -0.42              &     -0.42             \\
                 & CL2          &     0.18 (0.05)       &    0.16 (0.04)        &     0.08 (0.05)       \\
                 & TC           &     0.06 (0.04)       &    0.05 (0.05)        &    -0.04 (0.07)       \\
                 & LB           &     0.00 (0.01)       &    0.00 (0.01)        &     0.00 (0.01)       \\
                 & CLB          &     0.00 (0.00)       &    0.00 (0.00)        &     0.00 (0.00)       \\     \hline  
\end{tabular}
\end{center}
\end{table}   

\begin{table}[p]
\caption{\label{skewtresults3}
Methods L1, L2, CL1, CL2, TC, LB and CLB applied to integrate multivariate
skew t densities in $10$ dimensions with $3$ and $10$ degrees of freedom and
zero, moderate and extreme skewness. The estimated log of the integral of the
density (true value $0$) is reported. } 
\begin{center}
\begin{tabular}{c|c|ccc}
\multicolumn{5}{c}{} \\ \hline
Degrees of  & Method & \multicolumn{3}{c}{Skewness}  \\ \cline{3-5}
freedom       &               &       $\delta_1=0$  &         $\delta_1=0.5$ &       $\delta_1=0.99$ \\ \cline{3-5}
   3             & L1           &     -3.58             &    -3.58              &     -3.74             \\
                 & L2           &     3.89 (0.07)       &    3.72 (0.08)        &     2.89 (0.06)      \\
                 & CL1          &     -2.97             &    -2.97              &     -2.98             \\
                 & CL2          &     2.91 (0.08)       &    2.76 (0.08)        &     2.15 (0.08)       \\
                 & TC           &     0.16 (0.07)       &    0.03 (0.08)        &    -0.48 (0.34)       \\
                 & LB           &     0.00 (0.03)       &    0.00 (0.03)        &     0.00 (0.04)       \\
                 & CLB          &     0.00 (0.01)       &    0.00 (0.02)        &     0.01 (0.01)       \\     \hline
   10            & L1           &     -1.89             &    -1.89              &     -2.07             \\
                 & L2           &     1.51 (0.04)       &    1.48 (0.04)        &     1.19 (0.04)      \\
                 & CL1          &     -1.50             &    -1.50              &     -1.50             \\
                 & CL2          &     1.18 (0.09)       &    1.13 (0.08)        &     0.97 (0.06)       \\
                 & TC           &     0.10 (0.07)       &    0.08 (0.06)        &    -0.12 (0.05)       \\
                 & LB           &     0.00 (0.01)       &    0.00 (0.01)        &     0.00 (0.02)       \\
                 & CLB          &     0.00 (0.00)       &    0.00 (0.00)        &     0.00 (0.00)       \\     \hline  
\end{tabular}
\end{center}
\end{table}   

\begin{table}[p]
\caption{\label{birthwtpred}Predictors for low birth weights data set}
\begin{center}
\begin{tabular}{rl}
Predictor & \multicolumn{1}{c}{Description} \\ \hline
age & age of mother in years \\
lwt & weight of mother (lbs) at least menstrual period \\
raceblack & indicator for race=black (0/1) \\
raceother & indicator for race other than white or black (0/1) \\
smoke & smoking status during pregnancy (0/1) \\
ptd & previous premature labors (0/1) \\
ht & history of hypertension (0/1) \\
ui & has uterine irritability (0/1) \\
ftv1 & indicator for one physician visit in first trimester (0/1) \\
ftv2+ & indicator for two or more physician visits in first trimester (0/1) \\
\end{tabular}
\end{center}
\end{table}

\begin{table}[p]
\caption{\label{bwtresults1}
Approximations to
log marginal likelihoods for models $M_0$ (linear model with all original predictors
and no interactions), $M_1$ (interaction model of Venables and Ripley) and $M_2$ 
(model with additive terms for continuous covariates) for low birthweight example.   } 
\begin{center}
\begin{tabular}{c|c|ccc}
\multicolumn{5}{c}{} \\ \hline
Link function   & Method & \multicolumn{3}{c}{Model} \\ \hline
                     &               &       $M_0$  &          $M_1$ &        $M_2$    \\ \cline{3-5}
   Logistic       & L1           &     -124.3              &   -120.0          &    -122.4          \\
                  & L2           &     -124.4 (0.2)        &   -120.1 (0.3)          &    -122.5 (0.3)          \\ 
                  & CL1          &     -124.2              &   -119.8          &    -122.1          \\ 
                  & CL2          &     -124.3 (0.3)        &   -120.0 (0.5)          &    -122.4 (0.5)          \\ 
                  & TC           &     -124.6 (0.4)        &   -120.4 (0.4)          &  -123.1 (0.5)          \\ 
                  & LB           &     -124.1 (0.0)        &   -119.7 (0.0)          &    -122.0 (0.1)        \\                  
                  & CLB          &     -124.3 (0.0)        &   -120.0 (0.0)          &    -122.5 (0.1)          \\ 
                  & GS           &     -124.1              &   -119.7                &    -122.0          \\  \hline 
    Robit         & L1           &     -132.9              &   -128.1                &    -131.4          \\
                  & L2           &     -132.2 (0.2)        &   -127.2 (0.2)          &    -129.9 (0.3)          \\ 
                  & CL1          &     -132.7              &   -127.8                &    -129.6          \\ 
                  & CL2          &     -132.6 (0.3)        &   -127.8 (0.5)          &    -130.7 (0.5)          \\ 
                  & TC           &     -132.9 (0.4)        &   -127.9 (0.3)          &  -131.1 (0.5)                \\ 
                  & LB           &     -132.4 (0.0)        &   -127.4 (0.1)          &    -130.3 (0.1)          \\                  
                  & CLB          &     -132.7 (0.0)        &   -127.7 (0.0)          &    -130.8 (0.1)          \\ 
                  & GS           &     -132.5              &   -127.5                &    -130.3          \\ \hline
\end{tabular}
\end{center}
\end{table}

 \begin{table}[p]
\caption{\label{bwtresults2}
Approximations to
log marginal likelihoods for models $M_0$ (linear model with all original predictors
and no interactions), $M_1$ (interaction model of Venables and Ripley) and $M_2$ 
(model with additive terms for continuous covariates) for randomly chosen subset
of size 50 for low birthweight example.   } 
\begin{center}
\begin{tabular}{c|c|ccc}
\multicolumn{5}{c}{} \\ \hline
Link function   & Method & \multicolumn{3}{c}{Model} \\ \hline
                     &               &       $M_0$  &          $M_1$ &        $M_2$    \\ \cline{3-5}                 
   Logistic       & L1           &    -37.9         &     -38.1        &    -38.6          \\
                  & L2           &    -35.7 (0.4)   &     -34.8 (1.0)  &    -33.2 (0.6)          \\ 
                  & CL1          &    -36.9         &     -36.7        &    -35.7          \\ 
                  & CL2          &    -36.8 (0.3)   &     -36.5 (0.7)  &    -35.9 (0.4)          \\ 
                  & TC           &    -36.9 (0.3)   &     -36.6 (0.3)  &    -36.5 (0.9)           \\ 
                  & LB           &    -36.6 (0.2)   &     -36.0 (0.4)  &    -35.2 (0.6)          \\                  
                  & CLB          &    -36.9 (0.2)   &     -36.5 (0.1)  &    -36.0 (0.3)          \\ 
                  & GS           &    -36.6         &     -36.1        &    -35.4          \\  \hline 
    Robit         & L1           &    -43.2         &     -43.1        &    -43.1          \\
                  & L2           &    -39.3 (0.4)   &     -38.3 (0.9)  &    -36.8 (0.9)          \\ 
                  & CL1          &    -39.9         &     -38.5        &    -37.0          \\ 
                  & CL2          &    -41.1 (0.3)   &     -40.6 (0.6)  &    -39.8 (0.6)          \\ 
                  & TC           &    -41.5 (0.5)   &     -40.8 (0.8)  &    -40.1 (0.7)          \\ 
                  & LB           &    -40.8 (0.3)   &     -40.1 (0.7)  &    -39.0 (0.9)          \\                  
                  & CLB          &    -41.2 (0.4)   &     -40.6 (0.2)  &    -39.9 (0.4)          \\ 
                  & GS           &    -40.7         &     -40.3        &    -39.5          \\ \hline                         
\end{tabular}
\end{center}
\end{table}

\begin{table}[p]
\caption{\label{papsmear}Predictors for papsmear data set}
\begin{center}
\begin{tabular}{rl}
Predictor & \multicolumn{1}{c}{Description} \\ \hline
knowgp    & 1 if the GP is known to the patient; 0 otherwise \\
sexgp     & 1 if the GP is male; 0 otherwise \\
testdue   & 1 if the patient is due or overdue for a paptest; 0 otherwise \\
drrec     & 1 if the GP recommends that the patient has a paptest; 0 otherwise \\
papcost   & cost of test in Australian dollars \\
\end{tabular}
\end{center}
\end{table}

\begin{table}[p]
\caption{\label{CJresults}
Approximations to
log marginal likelihoods for heteroscedastic and homoscedastic models 
for the papsmear data.  }
\begin{center}
\begin{tabular}{c|cc}
\multicolumn{3}{c}{} \\ \hline
Method & \multicolumn{2}{c}{Model} \\ \hline
                     &   Heteroscedastic    &    Homoscedastic    \\ \cline{2-3}                 
L                     &  -1101.8 (0.5) & -1119.1 (0.2) \\
CL                    &  -1101.4 (0.5) & -1118.9 (0.2) \\
CJ                    &  -1101.5 (0.5) & -1118.9 (0.2) \\
\hline                         
\end{tabular}
\end{center}
\end{table}

\end{document}